\documentclass[aps,prd,twocolumn,groupedaddress,showpacs]{revtex4}

\usepackage{graphicx,dcolumn,bm,amssymb,amsmath,latexsym,amsfonts}

\begin{document}

\newcommand{\nonu}{\nonumber}
\newcommand{\sm}{\small}
\newcommand{\noi}{\noindent}
\newcommand{\npg}{\newpage}
\newcommand{\nl}{\newline}
\newcommand{\bc}{\begin{center}}
\newcommand{\ec}{\end{center}}
\newcommand{\be}{\begin{equation}}
\newcommand{\ee}{\end{equation}}
\newcommand{\beal}{\begin{align}}
\newcommand{\eeal}{\end{align}}
\newcommand{\bea}{\begin{eqnarray}}
\newcommand{\eea}{\end{eqnarray}}
\newcommand{\bnabla}{\mbox{\boldmath $\nabla$}}
\newcommand{\univec}{\textbf{a}}
\newcommand{\VectorA}{\textbf{A}}
\newcommand{\Pint}

\title{Binary system of unequal counterrotating Kerr-Newman sources}

\author{I. Cabrera-Munguia\footnote{icabreramunguia@gmail.com}}
\affiliation{Departamento de F\'isica, Universidad Aut\'onoma Metropolitana-Iztapalapa A.P. 55-534, M\'exico D.F. 09340, M\'exico \\ Departamento de Ciencias B\'asicas, Universidad Polit\'ecnica de L\'azaro C\'ardenas, 60950 L\'azaro C\'ardenas, Michoac\'an, M\'exico }


\begin{abstract}
Stationary axisymmetric binary systems of unequal counterrotating Kerr-Newman sources with a massless strut in between are studied. By means of the choice of a suitable parametrization, the axis conditions and the absence of individual magnetic charges are fulfilled; thus, the entire metric reduces to a 6-parametric asymptotically flat exact solution. Later on, with the purpose to describe interacting black holes, the analytic functional form of the horizon half-length parameter $\sigma_{k}$ is obtained explicitly in terms of physical Komar parameters: mass $M_{k}$, electric charge $Q_{k}$, angular momentum $J_{k}$, and coordinate distance $R$, where the seven physical parameters satisfy a simple algebraic relation. Finally, in the limit of extreme black holes, the full metric is derived in a closed analytical form, and a study on the absence or appearance of naked singularities off the axis is presented. \end{abstract}

\maketitle

\section{Introduction}
\vspace{-0.5cm}
In stationary axisymmetric spacetimes, the study of binary systems composed by Reissner-Norstr\"{o}m (RN) sources \cite{Reissner,Nordstrom} began to receive attention since the well-known Weyl family of solutions was presented in 1917 \cite{Weyl}; it relates the masses $M_{k}$ and electric charges $Q_{k}$, for $k=1,2$, through the relation $M_{1}Q_{2}-M_{2}Q_{1}=0$. In the context of black hole sources, a particular case of Weyl's family is the Majumdar-Papapetrou solution \cite{Majumdar,Papapetrou}, which describes two extreme black holes in equilibrium when the charges are equal to the masses accor\-ding to $Q_{k}=\pm M_{k}$, regardless of the distance $R$ between sources. Many years later, Varzugin and Chistyakov studied intensively these binary configurations in the black hole sector \cite{Varzugin1}, and, after solving the co\-rres\-pon\-ding Riemann-Hilbert problem, they found explicitly the formulas for the event horizons of length $ 2\sigma_{kE}$, which are given by
\vspace{-0.2cm}
\bea \begin{split}
& \qquad  \qquad \sigma_{1E}= \sqrt{M_{1}^{2}-Q_{1}^{2}+2\mu Q_{1}},\\
\sigma_{2E}&=\sqrt{M_{2}^{2}-Q_{2}^{2}-2\mu Q_{2}},\qquad
\mu:=\frac{M_{2}Q_{1}-M_{1}Q_{2}}{R+M_{1}+M_{2}}. \label{electrostaticsigma}\end{split}\eea

\vspace{-0.2cm}
The above set of Eqs.\ (\ref{electrostaticsigma}) is quite important to understand better the physical and thermodynamical pro\-per\-ties of such configurations, which were used in the years following its discovery. For instance, by applying soliton techniques the equilibrium problem without strut in bet\-ween the black holes is studied in Ref.\ \cite{Alekseev}, while the interaction force associated with the strut and the full metric are derived in Ref.\ \cite{MankoDRN}. The strut prevents the sources from falling onto each other; it represents a line source of pressure computed through the conical sin\-gu\-la\-ri\-ty (angle's deficit) \cite{BachWeyl,Israel}. The counterpart of the aforementioned charged solutions is the rotating case, which has been extensively studied since the double-Kerr-NUT solution was developed by Kramer and Neugebauer in 1980 \cite{KramerNeugebauer}, but nowadays in the vacuum case there e\-xist no analogous formulas like those given by Eqs.\ (\ref{electrostaticsigma}). It is worth mentioning, that charged or rotating binary systems develop naked singularities off the axis as singular surfaces (SS) \cite{IMR,RICM} or ring singularities (RS) \cite{DietzHoenselaers}, respectively, if at least one of the masses results to be negative \cite{MRS1,Hennig}.

On the other hand, the electrovacuum sector seems to be impossible to treat due to the technical difficulties provided by the electromagnetic field into rotating systems. Therefore, only equal cases have been s\-tu\-died after taking into account the advantages of their sy\-mme\-try properties in which the axis conditions are a\-u\-to\-ma\-ti\-ca\-lly fulfilled. A first study of a binary system of identical Kerr-Newman (KN) sources \cite{Newman} in equilibrium under their mutual electromagnetic and gravitational interactions was given by Parker, Rufinni, and Wilkins \cite{Parker} through the Perj\'es-Israel-Wilson method \cite{PIW1,PIW2}. The sources are two thin disks lying on the equatorial plane, whose charges are equal to their respective masses and contain opposite spin (a counterrotating system), and both charges are equal in magnitude and sign. Recently a more general description of such a problem is developed in Ref.\ \cite{MRR}, where the functional form of the horizon half-length parameter $\sigma$ is introduced in terms of physical Komar parameters \cite{Komar}. Furthermore, if the charges have opposite sign, the so\-lu\-tion represents a counterrotating black dihole system \cite{Emparan,EmparanTeo}, whose magnetic dipole moment is generated by the rotation of electrically charged black holes \cite{MRS,CLLM}. Additionally, it has been shown that individual magnetic charges can be created by the rotation of these diholes \cite{CLLM,ILLM}, and now the system turns out to be dyonic \cite{Schwinger}, where the Smarr mass formula \cite{Smarr} is generalized in order to include the contribution of the magnetic charge into the mass \cite{Tomimatsu}.

Since, in balancing configurations (without the su\-ppor\-ting strut) \cite{MRS1,Hennig,DietzHoenselaers}, the interpretation of such binary systems as describing two black holes is spoiled, this motivates the search on
stationary configurations of in\-te\-rac\-ting black hole sources with a strut in between, with the purpose to use this kind of solution and provide novel e\-vi\-den\-ce on the interactions and their mathematical pro\-per\-ties, which may be used to construct initial data in numerical simulations like the \emph{momentary stationary data} performed in Ref.\ \cite{Dain} for two extreme Kerr sources. The present paper aims at the construction of a model for un\-e\-qual counterrotating KN black hole sources interacting by means of a massless strut, in which the full metric and all its geometrical properties are given in a more phy\-si\-cal way through the derivation of $\sigma_{k}$ as a function of Komar's physical parameters.

To accomplish our goal, first we solve analytically the axis conditions in order to describe a binary system of unequal counterrotating KN sources with a su\-ppor\-ting strut. We will use the same idea given in Ref.\ \cite{MRS} to eliminate the individual magnetic charges as well as the total monopolar magnetic charge; thus, the Smarr formula for the mass is ensured. Afterward, we will be able to derive the nontrivial expressions for $\sigma_{k}$, $k=1,2$, which include the rotation parameter and generalize the aforementioned formulas given by Eq.\ (\ref{electrostaticsigma}). In fact, this paper deals with the unequal case of the solutions already discussed in Refs.\ \cite{MRR,MRS}, where now the seven physical parameters satisfy an algebraic relation which might be understood as a dynamic scenario between sources, since the physical properties of one body are a\-ffec\-ted by the presence of the other one. Later on, all the thermodynamical features of the system will be determined in a concise explicit form. The second objective pursued in this paper is the derivation of the extreme limit case in a closed analytical form, by means of Perj\'es' representation \cite{Perjes}. We will prove that the model saturates the in\-e\-qua\-li\-ty for interacting black holes with struts discovered by Gabach-Clement \cite{Maria}. Additionally, a numerical study on \emph{ring singularities} off the axis is given. In the extreme double-Reissner-Norstr\"{o}m (DRN) sector an easy a\-na\-ly\-ti\-cal proof is also provided on the absence or appearance of \emph{singular surfaces} as a complement of the well-known positive mass theorem \cite{SchoenYau1,SchoenYau2}.

\vspace{-0.7cm}
\section{The asymptotically flat exact solution}
\vspace{-0.5cm}
Let us start the section by introducing the Ernst equations \cite{Ernst} as follows:
\vspace{-0.2cm}
\bea \begin{split}  \left({\rm{Re}} {\cal{E}}+|\Phi|^{2}\right)\Delta{\cal{E}}&=(\bnabla{\cal{E}}+
2\bar{\Phi}\bnabla \Phi)\bnabla {\cal{E}}, \\
 \left({\rm{Re}}{\cal{E}}+|\Phi|^{2}\right)\Delta \Phi&=(\bnabla{\cal{E}}+
2\bar{\Phi}\bnabla\Phi)\bnabla\Phi, \label{ERNST} \end{split} \eea

\vspace{-0.2cm}
\noi which describe stationary axisymmetric electrovacuum spacetimes through the complex potentials ${\cal{E}}=f - |\Phi|^{2} + i\Psi$ and $\Phi=-A_{4} +i A_{3}^{'}$, where \bnabla and $\Delta$ are the gradient and Laplace operators, respectively, defined in Weyl-Papapetrou cylindrical coordinates $(\rho,z)$. The electrovacuum exact solution of Eq.\ (\ref{ERNST}) describing a binary system composed by KN sources, can be obtained by means of the Sibgatullin method (SM) \cite{Sibgatullin,RMJ} which is based on the soliton theory and is helpful to construct the Ernst potentials $({\cal{E}}, \Phi)$ in the whole spacetime as well as the metric functions $f(\rho,z)$, $\omega(\rho,z)$, and $\gamma(\rho,z)$ of the stationary axisymme\-tric line element \cite{Papapetrou0}
\be ds^{2}=f^{-1}\left[e^{2\gamma}(d\rho^{2}+dz^{2})+\rho^{2}d\varphi^{2}\right]- f(dt-\omega d\varphi)^{2}.
\label{Papapetrou}\ee

In this approach, the Ernst potentials on the symmetry axis ${\cal{E}}(\rho=0,z):=e(z)$ and $\Phi(\rho=0,z):=f(z)$ have the following form:
\vspace{-0.2cm}
\be e(z)= 1 + \sum_{j=1}^{2}\frac{e_{j}}{z-\beta_{j}}, \qquad
f(z)= \sum_{j=1}^{2}\frac{f_{j}}{z-\beta_{j}}, \label{Ernstsymmetryaxis}\ee

\begin{figure}[ht]
\centering
\includegraphics[width=6cm,height=5.0cm]{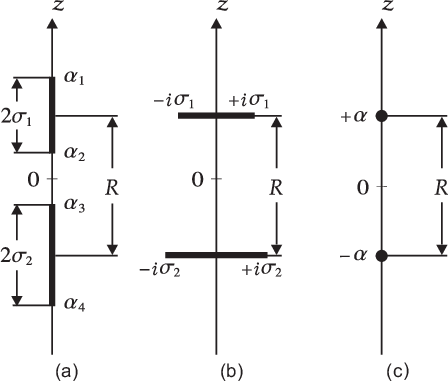}
\caption{Location of two unequal KN sources on the symmetry axis: (a) black hole configuration; (b) hyperextreme sources after making $\sigma_{k} \rightarrow i \sigma_{k}$; (c) the extreme limit case $\sigma_{k}=0$.}
\label{DKRR}\end{figure}

\noi where $\{e_{j},f_{j},\beta_{j}\}$ are complex constants which can be related with the Simon's multipolar moments \cite{Simon}. The specific choice (\ref{Ernstsymmetryaxis}) of the axis data is motivated by two facts: the mentioned relation of the parameters $e_{j}$, $f_{j}$, and $\beta_{j}$ with the multipolar terms and that the SM
deals with poles on the complex plane. Therefore, this specific representation turns out to work very well in the framework of the SM. With the purpose to describe a binary system of KN sources, the explicit solution is worked out after setting $N=2$ in the last formulas of Sec. III in \cite{RMJ}. Then, the two-body system is depicted by twelve algebraic parameters $\{\alpha_{n}, f_{j}, \beta_{j}\}$, for $n=\overline{1,4}$ and $j=1,2$, where $\alpha_{n}$ define the location of the sources on the symmetry axis (see Fig.\ \ref{DKRR}). It should be pointed out that SM provides a mathematical recipe to construct the Ernst potentials and metric functions at whole spacetime, but generally the solution is not asymptotically flat at spatial infinity, due to the presence of NUT sources \cite{NUT} and the global monopolar magnetic charge. Hence, to e\-li\-mi\-na\-te the gravitomagnetic monopole (NUT parameter) and disconnect the region between sources, one needs to solve the following axis conditions \cite{ILLM,FN}:
\bea  \begin{split}  {\rm{Im}}[\mathfrak{\bar{a}}_{-}(\mathfrak{g}_{-}+\mathfrak{h}_{-})]= 0, \qquad
{\rm{Im}}[\mathfrak{\bar{a}}_{+}(\mathfrak{g}_{+}+\mathfrak{h}_{+})]=0,&  \\
\mathfrak{g}_{\pm}=\left|
\begin{array}{ccccc}
0 & 2 & 2 & 1\pm1 & 1\pm1 \\
1 & {} & {} & {} & {}  \\
1 & {} & (\mathfrak{a}_{\pm})  \\
0 & {} & {} & {} & {} \\
0 & {} & {} & {} & {} \\
\end{array}
\right|,& \\
\mathfrak{h}_{\pm}=\left|
\begin{array}{ccccc}
0 & 1 & 1 & 1 & 1 \\
1 & {} & {} & {} & {}  \\
1 & {} & (\mathfrak{a}_{\pm})  \\
\bar{e}_{1} & {} & {} & {} & {} \\
\bar{e}_{2} & {} & {} & {} & {} \\
\end{array}
\right|,\\
\mathfrak{a}_{\pm}=\left|\begin{array}{cccc}
\pm \gamma_{11} & \pm \gamma_{12} & \gamma_{13}& \gamma_{14} \\
\pm \gamma_{21} & \pm \gamma_{22} & \gamma_{23}& \gamma_{24} \\
M_{11} & M_{12} & M_{13}& M_{14}\\
M_{21} & M_{22} & M_{23}& M_{24}\\
\end{array}
\right|,& \\
M_{jn}=\left[\bar{e}_{j}+2\bar{f}_{j} f(\alpha_{n})\right](\alpha_{n}-\bar{\beta}_{j})^{-1},& \\ f(\alpha_{n})=\sum_{j=1}^{2} f_{j}\gamma_{jn},\qquad
\gamma_{jn}=(\alpha_{n}-\beta_{j})^{-1},&\\
e_{1}=\frac{2 \prod_{n=1}^{4}(\beta_{1}-\alpha_{n})}
{(\beta_{1}-\beta_{2})(\beta_{1}-\bar{\beta}_{1})(\beta_{1}-\bar{\beta}_{2})}-\sum_{k=1}^{2} \frac{2f_{1}\bar{f}_{k}}{\beta_{1}-\bar{\beta}_{k}},&\\
e_{2}=\frac{2 \prod_{n=1}^{4}(\beta_{2}-\alpha_{n})}
{(\beta_{2}-\beta_{1})(\beta_{2}-\bar{\beta}_{1})(\beta_{2}-\bar{\beta}_{2})}-\sum_{k=1}^{2} \frac{2f_{2}\bar{f}_{k}}{\beta_{2}-\bar{\beta}_{k}}.&
\end{split} \label{algebraicequations}\eea

The parameters $\alpha_{n}$ can be written in terms of the re\-la\-ti\-ve distance $R$ and the half-length $\sigma_{k}$ of each rod as follows:
\vspace{-0.2cm}
\bea \begin{split}
\alpha_{1}&=\frac{R}{2}+\sigma_{1}, \qquad \alpha_{2}=\frac{R}{2}-\sigma_{1}, \\
\alpha_{3}&=-\frac{R}{2}+\sigma_{2},\qquad \alpha_{4}=-\frac{R}{2}-\sigma_{2}, \end{split} \eea

\noi where $\sigma_{k}$ can take real positive or pure imaginary values as shown in Fig.\ \ref{DKRR}. In order to solve the above set of algebraic equations (\ref{algebraicequations}), one notes that the first Simon's multipolar terms \cite{Simon} as the total mass $\mathcal{M}$, total electric charge $\mathcal{Q}$, and total magnetic charge $\mathcal{B}$ can be calculated from the Ernst potentials (\ref{Ernstsymmetryaxis}) on the symmetry axis, lea\-ding us to
\vspace{-0.2cm}
\be \beta_{1} + \beta_{2} + \bar{\beta}_{1} + \bar{\beta}_{2}=-2\mathcal{M}, \qquad f_{1}+f_{2}= \mathcal{Q}+ i\mathcal{B}. \label{Total}\ee

By choosing $\beta_{1} + \beta_{2}=-\mathcal{M}:=-M$ and $\mathcal{B}:=0$, one may describe a two-body system of counterrotating KN sources apart by a massless strut, where the upper cons\-ti\-tuent is endowed with mass $M_{1}$, electric and magnetic charge $Q_{1}$, $B_{1}$, respectively, and angular momentum $J_{1}$, while the lower one contains $M_{2}$, $Q_{2}$, $B_{2}$, and $J_{2}$, respectively. The individual magnetic charges are equal in magnitude but opposite in sign, i.e., $B_{1}=-B_{2}=Q_{B}$. The angular momenta have opposite spin, and both bodies are separated by the coordinate distance $R$. Moreover, if we want to remove also the individual magnetic charges $(B_{k}=0)$, the following condition should be satisfied \cite{Tomimatsu}:
\vspace{-0.4cm}
\bea \begin{split} A_{4}(\rho&=0,z=\alpha_{2k-1})-A_{4}(\rho=0,z=\alpha_{2k})=0,&\\
&k=1,2,  & \label{Tomimagnetic}\end{split} \eea
\vspace{-0.4cm}

\noi where $A_{4}$ is the electric potential computed from the real part of $\Phi$. Therefore, the explicit solution of the axis conditions (\ref{algebraicequations}), together with an absence of magnetic charges
Eq.\ (\ref{Tomimagnetic}) is given by
\vspace{-1.0cm}
\begin{widetext}
\bea \begin{split} f_{1,2}&=\pm \frac{Q \beta_{1,2}+ q_{o}+i b_{o}}{\beta_{1}-\beta_{2}},\qquad
\beta_{1,2}=\frac{-M\pm\sqrt{p+iq_{+}}}{2},\\
p&=R^{2}+M^{2}-2\Delta_{o}+2\left(\epsilon_{1}-\frac{\epsilon_{2}R}{M} \right)-\frac{4Qq_{o}}{M},\qquad
q_{\pm}=2\left(\frac{\beta_{o}(MR\mp \epsilon_{2}) \mp Q \beta}{M \alpha_{o}}\right)\delta, \qquad  b_{o}=\frac{\beta \delta}{2\alpha_{o}},\\
\delta&:=\sqrt{\alpha_{o}\left(\beta_{o}[M^{2}(\Delta_{o}-2\epsilon_{1})+\epsilon_{2}^{2}
+4q_{o}^{2}]-\beta^{2}\right)},\qquad
\alpha_{o}:=\beta_{o}\left[\beta_{o}(M^{2}R^{2}-\epsilon_{2}^{2})-2 \epsilon_{2}Q \beta\right]+\Delta_{o}\beta^{2},\\
\beta&:=2q_{o}R-\epsilon_{2}Q, \qquad \beta_{o}:=R^{2}-\Delta_{o},\qquad
\Delta_{o}:=M^{2}-Q^{2}, \qquad \epsilon_{1,2}:=\sigma_{1}^{2}\pm\sigma_{2}^{2}.
\label{implicitsolution2}\end{split}\eea

After cumbersome calculations the Ernst potentials and metric functions which define a 6-parametric asymptotically flat exact solution assume the form
\bea \begin{split}
{\cal{E}}&=\frac{\Lambda+\Gamma}{\Lambda-\Gamma},\quad \Phi=\frac{\chi}{\Lambda-\Gamma},\quad f=\frac{|\Lambda|^{2}-|\Gamma|^{2}+|\chi|^{2}}{| \Lambda-\Gamma|^{2}}, \quad \omega=\frac{{\rm{Im}}\left[(\Lambda-\Gamma)\bar{\mathcal{G}}-\chi\bar{\mathcal{I}}\right]}{|\Lambda|^{2}-
|\Gamma|^{2}+|\chi|^{2}},\quad e^{2\gamma}=\frac{|\Lambda|^{2}-|\Gamma|^{2}+|\chi|^{2}}{256\sigma_{1}^{2}\sigma_{2}^{2}\alpha_{o}^{2} r_{1}r_{2}r_{3}r_{4}}, \\
\Lambda&=4\sigma_{1}\sigma_{2}(\alpha_{o}-M^{2}\beta_{o}^{3})(r_{1}r_{2}+r_{3}r_{4})
+[2M^{2}\beta_{o}^{4}+(\Delta_{o}-\epsilon_{1}-\beta_{o})(\alpha_{o}-M^{2}\beta_{o}^{3})](r_{1}-r_{2})(r_{3}-r_{4})\\
&+2\sigma_{1}\sigma_{2}(\alpha_{o}+M^{2}\beta_{o}^{3})(r_{1}+r_{2})(r_{3}+r_{4})- 2 i \beta_{o}\delta [\sigma_{1}(r_{1}+r_{2})(r_{3}-r_{4})-\sigma_{2}(r_{1}-r_{2})(r_{3}+r_{4})],\\
\Gamma&=(2/MR)\left\{\sigma_{2}\left[2\sigma_{1}M^{2}(a_{o}\beta_{o}^{2}-\alpha_{o})(r_{1}+r_{2}) +\epsilon_{+}(r_{1}-r_{2})\right] -\sigma_{1}\left[2\sigma_{2}M^{2}(a_{o}\beta_{o}^{2}+\alpha_{o})(r_{3}+r_{4}) +\epsilon_{-}(r_{3}-r_{4})\right]\right\},\\
\chi&=(2/R)\left\{\sigma_{2}\left[2\sigma_{1}R( \alpha_{o}Q-M^{2}\beta\beta_{o}^{2})(r_{1}+r_{2}) +\varepsilon_{+}(r_{1}-r_{2})\right] +\sigma_{1}\left[2\sigma_{2}R( \alpha_{o}Q+M^{2}\beta\beta_{o}^{2})(r_{3}+r_{4}) +\varepsilon_{-}(r_{3}-r_{4})\right]\right\},\\
\mathcal{G}&=2z\Gamma + (1/R^{2})\{4\sigma_{1}\sigma_{2}\epsilon_{o} (r_{1}r_{2}-r_{3}r_{4})+\sigma_{1}\kappa_{-}(r_{1}+r_{2})(r_{3}-r_{4}) + \sigma_{2}\kappa_{+}(r_{1}-r_{2})(r_{3}+r_{4})-2i(a_{o}+\epsilon_{2}\beta) \\
&\times \delta R^{2}(r_{1}-r_{2})(r_{3}-r_{4})\} + (2/MR)\{\sigma_{2} [2\sigma_{1}R\nu_{+} (r_{1}+r_{2})+ \upsilon_{+} (r_{1}-r_{2})] + \sigma_{1} [2\sigma_{2}R\nu_{-} (r_{3}+r_{4})- \upsilon_{-} (r_{3}-r_{4})]\},\\
\mathcal{I}&= (1/MR^{2}) \{ 4\sigma_{1}\sigma_{2}R[M^{2}(a_{+}r_{1}r_{2}-a_{-}r_{3}r_{4})+2c_{o}R(\alpha_{o}+i\delta R)] + c_{o}R^{2}\left[c_{+}(r_{1}r_{4}+r_{2}r_{3})-c_{-}(r_{1}r_{3}+r_{2}r_{4})\right] +i\delta R^{2} \\
&\times\left[(c_{o}R-2M^{2}\beta)(r_{1}-r_{2})(r_{3}-r_{4})+ c_{o} \left[ (\sigma_{1}+\sigma_{2})(r_{1}r_{4}-r_{2}r_{3})-(\sigma_{1}-\sigma_{2})(r_{1}r_{3}-r_{2}r_{4}) \right]\right]
-\sigma_{1}d_{-}(r_{1}+r_{2})(r_{3}-r_{4})&\\
&+\sigma_{2}d_{+}(r_{1}-r_{2})(r_{3}+r_{4}) \} + (1/R)\{\sigma_{2}\left[2\sigma_{1} \lambda_{+}(r_{1}+r_{2})+ \mu_{+} (r_{1}- r_{2})\right]+\sigma_{1}\left[2\sigma_{2} \lambda_{-}(r_{3}+r_{4})+ \mu_{-} (r_{3}- r_{4})\right]\},\\
\epsilon_{\pm}&:= M^{2} (\alpha_{o}\Delta_{o}-\epsilon_{2}a_{o} \beta_{o}^{2}) \pm R(\alpha_{o} \nu_{o}
- a_{o} M^{2} \beta_{o}^{2}R)-i(M^{2}\beta_{o}\mp a_{o})\delta R, \qquad  a_{o}:=Q \beta+ \epsilon_{2} \beta_{o},\\
\varepsilon_{\pm}&:=M^{2}\beta \beta_{o}^{2}(R^{2}\pm \epsilon_{2})-\alpha_{o}(2q_{o}R \pm \Delta_{o}Q)-i(\beta \mp Q\beta_{o})\delta R, \qquad \epsilon_{o}:= 2(\alpha_{o}-M^{2}\beta_{o}^{2})R^{2}+\alpha_{o}Q^{2}-M^{2}\beta_{o}\beta^{2},\\
\kappa_{\pm}&:=2\alpha_{o}(\Delta_{o}+M^{2})R^{2}-\epsilon_{o}R^{2} \pm \left[ 2\epsilon_{2}(\alpha_{o}+M^{2}\beta_{o}^{3}+2M^{2}q_{o}^{2} \beta_{o})R^{2} +Q(2\beta+\epsilon_{2}Q)
(\alpha_{o}-\epsilon_{2}^{2}M^{2}\beta_{o}) \right],\\
\upsilon_{\pm}&:=\alpha_{o}[\nu_{o}\beta_{o} \pm M^{2}R(R^{2}-2\beta_{o}-2\epsilon_{1} \mp \epsilon_{2})] + M^{2}\beta_{o}[2\epsilon_{1}a_{o}\beta_{o}R -\epsilon_{2}\beta c_{o} -\beta_{o}(a_{o}\mp M^{2}\beta_{o})R^{3}]
+i\beta_{o}(\nu_{o} \mp M^{2}R)\delta R,\\
\nu_{\pm}&:= \alpha_{o} \nu_{o} \pm M^{4}\beta_{o}^{2^{}}R + i (a_{o} \pm M^{2}\beta_{o})\delta,  \qquad a_{\pm}:=\alpha_{o}Q-a_{o}\beta\beta_{o} \mp c_{o}\beta_{o}^{2}R,  \qquad c_{o}:=2\Delta_{o}q_{o}-\epsilon_{2}Q R \\
\mu_{\pm}&:=\alpha_{o}[Q(R^{2}+\beta_{o}-2\epsilon_{1} \mp \epsilon_{2})\pm \beta]\pm 2 c_{o}^{2} \beta  \pm M^{2}\beta_{o}^{2}[\beta(R^{2}-2\beta_{o}+2\epsilon_{1})\mp
2Q\beta_{o}(R^{2}\pm \epsilon_{2})\pm 3 \epsilon_{2}\beta ]\\
& -i [(Q R \pm4q_{o})\beta_{o}\mp 3R\beta]\delta R, \qquad \lambda_{\pm}:=\alpha_{o}(2q_{o}R\mp3Q \beta_{o}) \pm M^{2}\beta_{o}^{2}(2Q R^{2}\beta_{o}-3\epsilon_{2} \beta) +i(\beta \mp Q\beta_{o})\delta R, \\
c_{\pm}&:=\alpha_{o}-M^{2}\beta_{o}^{2}[R^{2}-(\sigma_{1} \pm \sigma_{2})^{2}], \qquad d_{\pm}:=\alpha_{o}[M^{2}R(2q_{o}\pm Q R)+a_{o}Q] \pm M^{2}a_{o}\beta \beta_{o}(R^{2}\pm\epsilon_{2}),
\label{ERNST0} \end{split} \eea
\end{widetext}

\noi where

\bea \begin{split} r_{1,2}&=\sqrt{\rho^{2}+\left(z-R/2 \mp \sigma_{1}\right)^{2}}, \\
r_{3,4}&=\sqrt{\rho^{2}+\left(z+R/2 \mp \sigma_{2}\right)^{2}}, \end{split}\eea

\noi and now the Ernst potentials on the symmetry axis given by Eq.\ (\ref{Ernstsymmetryaxis}) reduce to
\vspace{-0.2cm}
\bea \begin{split} e(z)&=\frac{e_{+}}{e_{-}}, \qquad f(z)=\frac{Q z + q_{o}+i b_{o}}{e_{-}}, \\
e_{\pm}&=z^{2} \mp M z + \frac{M(\Delta_{o}-\epsilon_{1}-R^{2}/2) \mp \nu_{o}}{2M} -i\frac{q_{\mp}}{4},\\
\nu_{o}&:=\epsilon_{2}R+2Q q_{o}. \end{split}\label{ernstsymmetry} \eea

At this point, it is worth mentioning that the procedure to eliminate the magnetic charges of the system was considered first in Ref.\ \cite{MRS} for a binary system of identical counterrotating KN black holes endowed with opposite electric charges, i.e., stationary black diholes. This particular solution \cite{MRS} has the advantage that naturally satisfies the axis conditions, due to its symmetry properties. Nevertheless, as was already mentioned in Ref.\ \cite{CLLM}, a more general procedure to determine new exact solutions has to do with the choice of a suitable parametrization in order to solve the axis conditions. In fact, the algebraic equations (\ref{algebraicequations}) represent a generalization of the axis conditions introduced in Ref.\ \cite{ICM} for vacuum solutions.

With respect to the total angular momentum of the system $J$, which can be obtained asymptotically from the Ernst potentials on the symmetry axis Eq.\ (\ref{ernstsymmetry}), it reads
\vspace{-0.3cm}
\be J=\frac{\epsilon_{2}(R^{2}-M^{2})+2Q q_{o}R}{2M \alpha_{o}}\delta,
\label{totalangularmomentum}\ee

\noi which under the transformation $\epsilon_{2}\rightarrow -\epsilon_{2}$, $z \rightarrow -z$, $q_{o} \rightarrow -q_{o}$ changes its global sign as well as the metric function $\omega$. This fact means that the metric function $\omega$ and the total angular momentum $J$ change their global signs if one exchanges the physical properties and the position of the constituents; i.e.,  $J_{(1\leftrightarrow 2)}=-J$ and $\omega(\rho,-z)_{(1\leftrightarrow 2)}=-\omega(\rho,z)$. Therefore, the full metric describes a two-body system of counterrotating KN sources. If $Q=0$ and $q_{o}=0$, Eq.\ (\ref{ERNST0}) is reduced to the vacuum solution presented in Ref.\ \cite{ICM}. Moreover, in the absence of rotation from Eq.\ (\ref{totalangularmomentum}) the following result is obtained:
\be q_{o}=\frac{\epsilon_{2}Q R \pm M \sqrt{(R^{2}-\Delta_{o})(\Delta_{o}^{2}-2\epsilon_{1}\Delta_{o} +\epsilon_{2}^{2})}}{2\Delta_{o}},\label{dipole}\ee

\noi where now Eq.\ (\ref{ERNST0}) defines a binary system of RN sources. The case concerning black holes \cite{Alekseev,MankoDRN} is straightforwardly obtained after using the Varzugin-Chistyakov parametrization  \cite{Varzugin1}, which is described by Eqs.\ (\ref{electrostaticsigma}). Ho\-we\-ver, the reduced metric given by
Eq.\ (\ref{dipole}) can be useful also to describe relativistic disks (hyperextreme sources) under the transformation $\sigma_{k} \rightarrow  i \sigma_{k}$. In what follows in this paper, we are interested only in the description of a two-body system of unequal counterrotating KN black holes, by means of an explicit derivation of the formulas for the two horizons $\sigma_{k}$ as a function of physical Komar parameters.

\vspace{-0.4cm}
\subsection{Physical representation for $\sigma_{k}$}
\vspace{-0.4cm}
As was already pointed out, $\sigma_{k}$ defines the half-length of the rod representing the $k$th black hole,
whose event horizon is defined as a null hypersurface $H_{k}=\{\alpha_{2k}\leq z \leq \alpha_{2k-1}, 0\leq \varphi \leq 2\pi, \rho\rightarrow 0\}$. A starting point for the physical representation is to use the well-known Tomimatsu's formulas \cite{Tomimatsu},
\vspace{-0.2cm}
\bea \begin{split} M_{k}&= -\frac{1}{8\pi}\int_{H_{k}} \omega \Psi_{,z}\, d\varphi dz, \quad
Q_{k}=\frac{1}{4\pi}\int_{H_{k}}\omega A_{3,z}^{'}\, d\varphi dz, \\
J_{k}&=-\frac{1}{8\pi}\int_{H_{k}}\omega\left[1+ \frac{\omega \Psi_{,z}}{2}
-\tilde{A}_{3}A_{3,z}^{'}-(A_{3}^{'}A_{3})_{,z}\right]d\varphi dz ,\\ \end{split}\label{Tomy}\eea

\noi with $\tilde{A}_{3}:=A_{3}+ \omega A_{4}$ and $\Psi={\rm Im}(\cal{E})$. The magnetic potential $A_{3}$ is obtained after taking the real part of Ki\-nners\-ley's potential $\Phi_{2}$ \cite{Kinnersley}, whose expression is performed within the framework of SM \cite{RMJ} and has the form
\vspace{-0.2cm}
\be A_{3}={\rm Re}\left(\Phi_{2}\right)={\rm Re}\left(-i\frac{I}{E_{-}}\right)=-z A_{3}^{'} + {\rm Im} \left(  \frac{\mathcal{I}}{\Lambda-\Gamma}\right).\ee

The corresponding masses $M_{k}$ and electric charges $Q_{k}$ are given, respectively, by
\vspace{-0.2cm}
\bea \begin{split}
M_{1,2}&=\frac{M}{2}\pm \frac{2Qq_{o}R+ \epsilon_{2}(R^{2}-M^{2})}{2M(R^{2}-M^{2}+Q^{2})},\\
Q_{1,2}&=\frac{Q}{2}\pm \frac{2q_{o}(MR+M^{2}-Q^{2})-\epsilon_{2}Q(R+M)}{2M(R^{2}-M^{2}+Q^{2})},\label{massesandcharges}\end{split}\eea

\noi and it can be shown from Eqs.\ (\ref{massesandcharges}) that $M=M_{1}+M_{2}$ and $Q=Q_{1}+Q_{2}$. Additionally, the following relations arise:
\vspace{-0.2cm}
\bea \begin{split}
q_{o}&=\frac{1}{2}Q_{1}(R-2M_{2})-\frac{1}{2}Q_{2}(R-2M_{1}),\\
\epsilon_{2}&=\sigma^{2}_{1E}-
\sigma^{2}_{2E},\label{relations}\end{split}\eea

\noi where $\sigma_{1E}$ and $\sigma_{2E}$ are the expressions of Eq.\ (\ref{electrostaticsigma}) for the case of two electrostatic black hole horizons co\-rres\-pon\-ding to the DRN problem, given in Ref.\ \cite{Varzugin1}. Because of the fact that there are no individual magnetic charges, the last term which defines the angular momentum $J_{k}$ does not make any contribution to the individual mass $M_{k}$; therefore, the Smarr formula for the mass \cite{Smarr} is fulfilled:
\bea \begin{split}
 M_{k}&=\frac{\kappa_{k}S_{k}}{4\pi} + 2\Omega_{k}J_{k}+ \Phi_{k}^{H}Q_{k}\\
&= \sigma_{k} + 2\Omega_{k}J_{k} + \Phi^{H}_{k}Q_{k}, \qquad k=1,2, \label{Smarrformula}\end{split}\eea

\noi where $\Phi^{H}_{k}=-A_{4_{k}}^{H}-\Omega_{k} A_{3_{k}}^{H}$ is the electric potential measured over the black hole horizon $H_{k}$, $\Omega_{k}:=1/\omega^{H_{k}}$ is the angular velocity, and $\omega^{H_{k}}$ is the metric function $\omega$ e\-va\-lua\-ted at the horizon. Straightforward calculations lead us to the following expressions for $\Omega_{k}$ and $\Phi^{H}_{k}$:
\begin{widetext}
\bea \begin{split}
\Omega_{k}&=\frac{ \sqrt{(R^{2}-M^{2}+Q^{2})(\sigma_{kE}^{2}-\sigma_{k}^{2})P_{0}}}{(R+M)[2P_{1k}(M_{k}+\sigma_{k})-Q_{k}P_{2k}]},\qquad
\Phi^{H}_{k}=\frac{(R+M)P_{2k}(M_{k}+\sigma_{k})-2Q_{k}P_{1}}{(R+M)[2P_{1k}(M_{k}+\sigma_{k})-Q_{k}P_{2k}]},\qquad k=1,2,\\
P_{0}&:=\left[(R+M_{1})^{2}-M_{2}^{2}\right]\left[(R+M_{2})^{2}-M_{1}^{2}\right]+[(Q_{1}-Q_{2})R+(M_{1}-M_{2})Q]^{2},\\
P_{1}&:=[Q_{1}(R+M_{1}-M_{2})+M_{1}Q_{2}][Q_{2}(R-M_{1}+M_{2})+M_{2}Q_{1}]-M_{1}M_{2}Q_{1}Q_{2},\\
P_{11}&:=M_{1}\left[(R+M_{2})^{2}-M_{1}^{2}\right]-Q_{1}\left[ Q_{2}R-(M_{1}-M_{2})Q \right],\qquad
P_{12}=P_{11 (1\leftrightarrow2)},\\
P_{21}&:=2(M_{1}-M_{2})(M_{2}Q_{1}-M_{1}Q_{2})+R(Q_{1}R+2M_{1}Q_{2})-Q_{1}(\sigma_{1E}^{2}-\sigma_{2E}^{2}),\qquad
P_{22}=P_{21 (1\leftrightarrow2)}.\label{newgeometrical} \end{split}\eea

By placing these last expressions in the right-hand side of the Smarr mass formula Eq.\ (\ref{Smarrformula}), one obtains $\sigma_{k}$ as a function of the physical Komar parameters and the coordinate distance
\bea \begin{split}
\sigma_{1}&= \sqrt{M_{1}^{2}-Q_{1}^{2}+2\mu Q_{1}-
\frac{J_{1}^{2}(R^{2}-\Delta_{o})\left(\left[(R+M_{1})^{2}-M_{2}^{2}\right]\left[(R+M_{2})^{2}-M_{1}^{2}\right]+
[(Q_{1}-Q_{2})R+(M_{1}-M_{2})Q]^{2}\right)}{(R+M)^{2}\left(M_{1}\left[(R+M_{2})^{2}-M_{1}^{2}\right]
-Q_{1}\left[ Q_{2}R-(M_{1}-M_{2})Q \right]\right)^{2}}},&\\
\sigma_{2}&= \sqrt{M_{2}^{2}-Q_{2}^{2}-2\mu Q_{2}-
\frac{J_{2}^{2}(R^{2}-\Delta_{o})\left(\left[(R+M_{1})^{2}-M_{2}^{2}\right]\left[(R+M_{2})^{2}-M_{1}^{2}\right]+
[(Q_{1}-Q_{2})R+(M_{1}-M_{2})Q]^{2}\right)}{(R+M)^{2}\left(M_{2}\left[(R+M_{1})^{2}-M_{2}^{2}\right]
-Q_{2}\left[ Q_{1}R+(M_{1}-M_{2})Q \right]\right)^{2}}},&\label{sigmas} \end{split}\eea

\noi whereas the second equation of (\ref{relations}) implies the following relation between the seven physical parameters:
\bea \begin{split}
&M_{1}M_{2}(R+M_{1}+M_{2})\left[J_{1} + J_{2}+ R\left(\frac{J_{1}}{M_{1}}+ \frac{J_{2}}{M_{2}} \right)- M_{1}M_{2}
\left(\frac{J_{1}}{M_{1}^{2}}+\frac{J_{2}}{M_{2}^{2}}\right)\right]\\
&+(M_{1}-M_{2})(Q_{1}+Q_{2})(Q_{1}J_{2}-Q_{2}J_{1})-Q_{1}Q_{2}(J_{1}+J_{2})R=0.
\label{relationmomentum} \end{split}\eea
\end{widetext}

For completeness, it can be shown that the total angular momentum of the system defined by
Eq.\ (\ref{totalangularmomentum}) accounts exactly for the sum of the individual angular momenta, i.e.,
\be J= J_{1}+J_{2}.\label{totalmomentum}\ee

An interesting physical property of this kind of con\-fi\-gu\-ra\-tion is the interaction force associated with the strut in between the black holes \cite{Israel}, which can be calculated via the formula \cite{Israel,Weinstein}
\bea \begin{split}
\mathcal{F}&=\frac{1}{4}(e^{-\gamma_{0}}-1)=\frac{M_{1}M_{2}-(Q_{1}-\mu)(Q_{2}+\mu)}{R^{2}-(M_{1}+M_{2})^{2}
+(Q_{1}+Q_{2})^{2}},\label{force} \end{split}\eea

\noi where $\gamma_{0}$ is the value of the metric function $\gamma$ on the region of the strut. One should observe that formula Eq.\ (\ref{force}) shows no explicit dependence of the angular momentum into this force. Indeed,
this formula was derived in Ref.\ \cite{MankoDRN}, and it corresponds to the DRN problem. Nonetheless, the contribution of the angular momentum arises from Eq.\ (\ref{relationmomentum}), which relates the seven physical parameters.

\vspace{-0.8cm}\subsection{Physical and geometrical properties of the solution}
\vspace{-0.4cm}
The thermodynamical characteristics of the solution are contained in the Smarr mass formula Eq.\ (\ref{Smarrformula}), in which the surface gravity $\kappa_{k}$ and area of the horizon $S_{k}$ of the $k$th black hole, are related to each other by means of $\sigma_{k}$.  The area of the horizon $S_{k}$ is calculated via the formulas \cite{Tomimatsu,Carter}
\be S_{k}=\frac{4\pi \sigma_{k}}{\kappa_{k}}, \qquad \kappa_{k}=\sqrt{-\Omega_{k}^{2}e^{-2\gamma^{H_{k}}}},\ee

\noi where $\gamma^{H_{k}}$ is the metric function $\gamma$ evaluated at the co\-rres\-pon\-ding horizon $H_{k}$. The area of the horizons, surface gravities, and angular velocities acquire the final form
\vspace{-0.9cm}
\begin{widetext}
\bea \begin{split}
S_{1}&= \frac{4\pi(R+M)^{2} [2P_{11}(M_{1}+\sigma_{1})-Q_{1}P_{21}]}{P_{0}}= 4\pi\frac{P_{11}^{2}\left[(R+M)(M_{1}+\sigma_{1})-Q_{1}Q\right]^{2}+ \left[J_{1}(R-M_{1}+M_{2})(R^{2}-\Delta_{o})\right]^{2}}{P_{11}^{2}\left[(R+\sigma_{1})^{2}-\sigma_{2}^{2}\right]},\\
\kappa_{1}&=\frac{\sigma_{1}P_{0}}{(R+M)^{2}[2P_{11}(M_{1}+\sigma_{1})-Q_{1}P_{21}]}= \frac{\sigma_{1}P_{11}^{2}\left[(R+\sigma_{1})^{2}-\sigma_{2}^{2}\right]}{P_{11}^{2}
\left[(R+M)(M_{1}+\sigma_{1})-Q_{1}Q\right]^{2}+ \left[J_{1}(R-M_{1}+M_{2})(R^{2}-\Delta_{o})\right]^{2}},\\
\Omega_{1}&=\frac{J_{1}(R^{2}-\Delta_{o})P_{o}}{(R+M)^{2}P_{11}[2P_{11}(M_{1}+\sigma_{1})-Q_{1}P_{21}]}
=\frac{P_{11}\left[(R+\sigma_{1})^{2}-\sigma_{2}^{2}\right]J_{1}(R^{2}-\Delta_{o})}{P_{11}^{2}
\left[(R+M)(M_{1}+\sigma_{1})-Q_{1}Q\right]^{2}+ \left[J_{1}(R-M_{1}+M_{2})(R^{2}-\Delta_{o})\right]^{2}},\\
S_{2}&=S_{1 (1\leftrightarrow 2)}, \qquad \kappa_{2}=\kappa_{1 (1\leftrightarrow 2)}, \qquad \Omega_{2}=\Omega_{1 (1\leftrightarrow 2)}. \label{Horizonproperties}\end{split}\eea
\end{widetext}

The above formulas (\ref{Horizonproperties}) are expressed in two e\-qui\-va\-lent forms, where the left-hand side of $S_{k}$ and $\kappa_{k}$ can be obtained directly after combining Eq.\ (\ref{ERNST0}) with
Eq.\ (\ref{massesandcharges}), and without any previous knowledge of the explicit form of $\sigma_{k}$. The electric and magnetic dipole moments expressed in physical parameters are reduced to
\bea \begin{split}
q_{o}&=\frac{1}{2}Q_{1}(R-2M_{2})-\frac{1}{2}Q_{2}(R-2M_{1}),\\
b_{o}&=\left[\frac{Q_{1}J_{1}(R+M_{1}-M_{2})}{P_{11}}+\frac{Q_{2}J_{2}(R-M_{1}+M_{2})}{P_{12}}\right]&\\
 &\times \left(\frac{R^{2}-\Delta_{o}}{R+M_{1}+M_{2}}\right),\label{dipoles}\end{split}\eea

\noi and it follows that our solution derives the identical cases already discussed in Refs.\ \cite{MRR, MRS}. For instance, the Bret\'on-Manko solution \cite{MRR} is obtained after setting $M_{1}=M_{2}=m$, $Q_{1}=Q_{2}=q$, and $J_{1}=-J_{2}=j$, where the unique $\sigma$ reads
\be \sigma= \sqrt{m^{2}-q^{2}-\frac{j^{2} (R^{2}-4m^{2}+4q^{2})}{[m(R+2m)-q^{2}]^{2}}}.\ee

Moreover, if $M_{1}=M_{2}=m$, $Q_{1}=-Q_{2}=q$, and $J_{1}=-J_{2}=j$, the corresponding solution for stationary black diholes \cite{MRS} is given by
\be \sigma= \sqrt{m^{2}-\left[q^{2}+\frac{j^{2}[(R+2m)^{2}+4q^{2}]}
{[m(R+2m)+q^{2}]^{2}}\right]\frac{R-2m}{R+2m}}.\label{stationaryBDH}\ee

In both cases, Eq.\ (\ref{relationmomentum}) is satisfied automatically. It should be noticed that in the limit $R\rightarrow \infty$ the strut in between vanishes and formulas (\ref{sigmas}) reduce to $\sigma_{k}=\sqrt{M_{k}^{2}-Q_{k}^{2}-J_{k}^{2}/M_{k}^{2}}$, therefore, the uniqueness theorem holds for the isolated KN black hole \cite{Carter}. The individual magnetic charges $B_{k}$ could be included after applying a \emph{duality rotation}, as was performed in Ref.\ \cite{MRS} for stationary black diholes.

On the other hand, magnetic charges of equal magnitude but opposite sign could be included from the beginning of the construction of the solution, in order to eliminate the global monopolar magnetic charge of the system $(\mathcal{B}=0)$. As Tomimatsu proposed \cite{Tomimatsu}, this me\-cha\-nism will include contributions from the Dirac string to the electromagnetic part of the individual angular momentum; thus, the Smarr formula for the mass does not hold \cite{CLLM}. Naturally, this violates the uniqueness theorem when sources are far away from each other, due to the fact that in the limit $R\rightarrow \infty$ one should recover the horizon for one isolated KN black hole joined to a monopolar magnetic charge \cite{CLLM}; i.e., the isolated KN black hole has a monopole hair.

Turning now to our binary system, if $R\rightarrow \sqrt{M^{2}-Q^{2}}$ the angular velocities stop, since both horizons are tou\-ching each other. For such a situation, the interaction force given by Eq.\ (\ref{force}) tends its value to infinity, and the system collapses and evolves as one single RN black hole \cite{Reissner,Nordstrom} of mass $M$ and charge $Q$.

\vspace{0.5cm}\section{The extreme limit case and some geometrical properties}
\vspace{-0.6cm}
The extreme limit case arises as a 4-parametric exact solution after setting $\sigma_{k}=0$ in Eq.\ (\ref{ERNST0}). In that case, a simple representation for the metric functions $f$, $\omega$, and $\gamma$ is written down in terms of four basic polynomials $\mu_{o}$, $\sigma_{o}$, $\pi_{o}$, and $\tau_{o}$:
\vspace{-0.5cm}\begin{widetext}
\bea \begin{split} {\cal{E}}&=\frac{A-B}{A+B},\qquad \Phi=\frac{F}{A+B}, \qquad \Phi_{2}=\frac{-i I}{A+B},\qquad f=\frac{D}{N}, \qquad
\omega=\frac{R(y^{2}-1)W}{2D}, \qquad
e^{2\gamma}=\frac{D}{M^{4}R^{8}(x^{2}-y^{2})^{4}}, \\
A&= M^{2}R^{2}[\beta_{o}(x^{2}-y^{2})^{2}+ \Delta_{o}(x^{4}-1)]
+ 4\Delta_{o}(q_{o}^{2} + b_{o}^{2})(1-y^{4})+2i \delta_{o}M^{2} R^{2}(x^{2}+y^{2}-2 x^{2} y^{2}),\\
B&=2M[M^{2}R x\Gamma_{+}+ 2Q(q_{o}+ i b_{o})y\Gamma_{-}],\qquad
F=2M^{2}[Q R x\Gamma_{+}+ 2(q_{o}+ i b_{o})y\Gamma_{-}],\\
I&=MR^{2}Q \left\{ M y \left[ \Gamma_{+}+2(MRx+\Delta_{o}-i\delta_{o})\right] +4Q(q_{o}y^{2}+ib_{o})\right\}(1-x^{2})+ 2(q_{o}+ib_{o})(MRx+\Delta_{o})\\
& \times \left\{M \left[ \Gamma_{-}+2(MRx+\Delta_{o}+i\delta_{o})\right] +4Q(q_{o}-ib_{o})y\right\}(1-y^{2}),\\
\Gamma_{\pm}&= \left(\sqrt{\Delta_{o}-b}\mp i\sqrt{\beta_{o}}\right)
 \left[\sqrt{\Delta_{o}-b}(x^{2}-1) \pm i\sqrt{\beta_{o}}(x^{2}-y^{2})\right]
+ b(x^{2}-1), \\
D&=\mu_{o}^{2}+(x^{2}-1)(y^{2}-1)\sigma_{o}^{2},\qquad
N=D+ \mu_{o} \pi_{o}-(1-y^{2})\sigma_{o}\tau_{o}, \qquad W=(x^{2}-1)\sigma_{o} \pi_{o}-\mu_{o}\tau_{o},\\
\mu_{o}&=M^{2}R^{2}[\beta_{o}(x^{2}-y^{2})^{2}+\Delta_{o}(x^{2}-1)^{2}]
-4\Delta_{o}(q_{o}^{2} + b_{o}^{2})(y^{2}-1)^{2},\qquad
\sigma_{o}=4 M^{2}R^{2}\delta_{o} x y,\\
\pi_{o}&=(4/M) \{ M^{2}R \left( M x [(Rx+M)(M R x+\Delta_{o})-M\beta_{o}y^{2}]+2Qq_{o}y[R(x^{2}-y^{2})+2Mx]\right)\\
&+4(q_{o}^{2}+b_{o}^{2})y[M(M^{2}-2\Delta_{o})y+2Q\Delta_{o}(q_{o}/\beta_{o})(1+y^{2})]  \},\qquad
\tau_{o}=4MR\delta_{o}(x^{2}-1)[M^{2}y-2Q (q_{o}/\beta_{o})(Rx+M)],\\
\delta_{o}&:=\frac{\sqrt{\Delta_{o}[M^{2}\beta_{o}-4(q_{o}^{2}+b_{o}^{2})]}}{M}, \qquad b:=\frac{4\Delta_{o}(q_{o}^{2}+b_{o}^{2})}{M^{2}\beta_{o}},
\qquad b_{o}:=q_{o} \sqrt{\frac{\Delta_{o}(M^{2}\beta_{o}-4 q_{o}^{2})}{M^{2}\beta_{o}^{2}+4\Delta_{o}q_{o}^{2}}}, \label{extreme}\end{split}\eea
\end{widetext}

\noi where $(x,y)$ are prolate spheroidal coordinates defined as
\bea \begin{split} x&=\frac{r_{+}+r_{-}}{2\alpha}, \qquad y=\frac{r_{+}-r_{-}}{2\alpha}, \\
 r_{\pm}&=\sqrt{\rho^{2} +(z\pm\alpha)^{2}}, \qquad \alpha:=\frac{R}{2}, \label{prolates}\end{split}\eea

\noi and thereby one gets a simple representation for the metric functions which is analogous to the one proposed by Perj\'es \cite{Perjes} in relation to the well-known Tomimatsu-Sato spacetimes \cite{TS}. The above Eq.\ (\ref{extreme}) shows also the Kinnersley potential $\Phi_{2}$ \cite{Kinnersley} with the purpose to get straightforwardly the magnetic potential $A_{3}$. The individual angular momentum $J_{k}$ is obtained whether the extremality condition is achieved, whose expression yields from Eq.\ (\ref{sigmas})
\vspace{-0.2cm}
\be J_{k}= \varepsilon_{k}\frac{\sigma_{kE}P_{1k}(R+M)}{\sqrt{P_{0}(R^{2}-\Delta_{o})}}, \qquad k=1,2, \ee

\noi where $\varepsilon_{1}=-\varepsilon_{2}=\varepsilon=\pm 1$. The masses $M_{k}$ and charges $Q_{k}$ are related by means of the second formula of Eq.\ (\ref{relations}), as follows:
\be \epsilon_{2}:=M_{1}^{2}-Q_{1}^{2}+2\mu Q_{1} - \left(M_{2}^{2}-Q_{2}^{2}-2\mu Q_{2} \right)=0. \label{relation}\ee

\noi On the other hand, Gabach Clement has recently dis\-co\-ve\-red \cite{Maria} a geometrical inequality for black holes with struts, which is given by
\be  \sqrt{1+4\mathcal{F}}\geq \frac{\sqrt{(8\pi J_{k})^{2}+ (4\pi Q_{k}^{2})^{2}}}{S_{k}}, \qquad k=1,2, \ee

\noi and, therefore, it can be shown that the binary black hole system saturates the aforementioned inequality; i.e., the equality sign is reached in the extreme limit case pre\-vious\-ly defined by Eq.\ ({\ref{extreme}}). In order to prove the equality into the geometrical inequality between extreme black holes with struts, we have that the area of the horizon is now reduced to
\bea \begin{split} S_{k}&= \frac{4\pi(R+M)^{2}(2M_{k}P_{1k}-Q_{k}P_{2k})}{P_{0}}, \qquad k=1,2.  \end{split} \eea

Since the interaction force $\mathcal{F}$ does not show explicit dependence of the angular momentum, the binary system saturates the inequality, namely,
\bea \begin{split} \sqrt{1+4\mathcal{F}}&= \frac{\sqrt{(8\pi J_{k})^{2}+ (4\pi Q_{k}^{2})^{2}}}{S_{k}}, \qquad k=1,2.  \end{split} \eea

\vspace{-0.9cm}\section{Singularities off the axis}
\vspace{-0.3cm}
\subsection{Ring singularities}
\vspace{-0.4cm}
According to the well-known positive mass theorem \cite{SchoenYau1,SchoenYau2}, a regular solution which is free of singularities allows only positive values for the total Arnowitt-Deser-Misner (ADM) mass of the system \cite{ADM}. However, the theorem is not enough proof to provide regularity in the solution, since a positive ADM mass given by the inequa\-lity $M=M_{1}+M_{2}>0$ can be ensured even if one of the masses is negative. Therefore, one should add analytic conditions that guarantee the regularity of the solution. In order to analyze whether the solution Eq.\ (\ref{extreme}) is regular outside the axis, we need to look at the denominator of the Ernst potentials, where a singular solution occurs if
\vspace{-0.2cm}
\be A+B= F_{R}+i F_{I}=0,\ee

\vspace{-0.4cm}
\noi for which
\vspace{-0.2cm}
\begin{widetext}
\bea \begin{split} F_{R}&= M^{2}R^{2}[(R^{2}-\Delta_{o})(x^{2}-y^{2})^{2}+ \Delta_{o}(x^{4}-1)]+ 4\Delta_{o}(q_{o}^{2} + b_{o}^{2})(1-y^{4})\\
&+ 2M \left\{ (M^{2}R x + 2Q q_{o}y)\left[(R^{2}-\Delta_{o})(x^{2}-y^{2})
+\Delta_{o}(x^{2}-1)\right]+2Q \delta_{o} b_{o}y(1-y^{2}) \right\}=0,\\
F_{I}&=2\delta_{o}M^{2}R^{2}(x^{2}+y^{2}-2x^{2}y^{2})+4M Q R^{2}b_{o}y(x^{2}-y^{2})+2M(1-y^{2})\left[\delta_{o}(M^{2}R x-2Qq_{o}y)-2Q\Delta_{o}b_{o}y\right]=0. \label{polynomials} \end{split}\eea
\end{widetext}

\vspace{-0.2cm}
Any solution of Eq.\ (\ref{polynomials}) represents a single point outside the symmetry axis named \emph{ring singularity}. However, due to the high order polynomials of Eq.\ (\ref{polynomials}) it is not po\-ssi\-ble to concrete an analytical study on the appearance or absence of naked singularities; therefore, one must resort to numerical analysis. With regard to Eq.\ (\ref{prolates}), now the cylindrical coordinates $(\rho,z)$ are given by
\vspace{-0.2cm}
\be \rho=\alpha\sqrt{(x^{2}-1)(1-y^{2})}, \qquad z=\alpha xy. \label{cilindricas} \ee

\noi and it shows that the region $x>1$, $|y|<1$ defines the va\-lues that Eq.\ (\ref{extreme}) can take in the plane $(x,y)$ to develop RS. After assigning a wide range of numerical values for the parameters of the solution, the curves depicted by Eq.\ (\ref{polynomials}) present no intersections in such a region if both masses are positive. Furthermore, if at least one of the masses is negative, an intersection occurs in this region as a consequence of the presence of a negative mass. Table \ref{table1} shows a set of numerical values for the physical pa\-ra\-me\-ters of the solution, Eq.\ (\ref{extreme}).
\vspace{-0.4cm}
\begin{table}[ht]
\centering
\caption{Numerical values for the extreme limit case of the double-Kerr-Newman problem.}
\begin{tabular}{c c c c c c c c  }
\hline \hline
$M_{1}$&$ M_{2}$&$Q_{1}$&$Q_{2}$&$J_{1}$&$J_{2}$&$R$ \\ \hline
  1.5  &   1    &  1.4  & 0.46  & -2.275& 1.951 & 2.8    \\
  1    &   0.5  &  1.1  & -0.35 & 0.886 & -0.67 & 2.5    \\
  1    &   2    &  0.5  & 2.143 & 7.574 & -7.789& 1.5 \\
  1    &  -0.6  &  0.7  & -0.1  & 0.306 &  0.855&  2 \\
  -1    & -1.1  &  0.1  & -0.191& -0.148& 0.179 & 2.4  \\
  \hline \hline
\end{tabular}
\label{table1}
\end{table}

\begin{figure}[ht]
\begin{minipage}{0.49\linewidth}
\centering
\includegraphics[width=4.25cm,height=5cm]{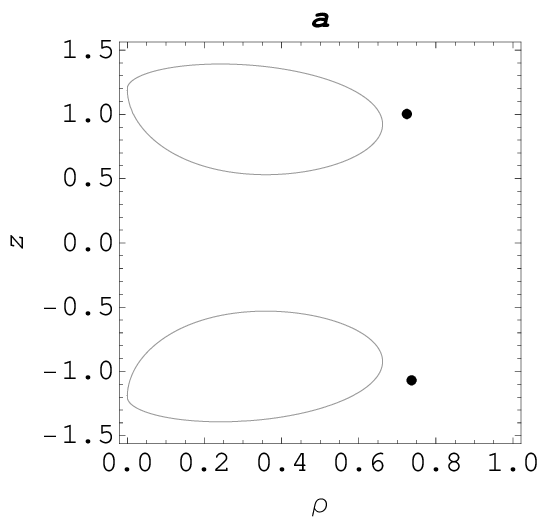}
\end{minipage}
\begin{minipage}{0.49\linewidth}
\centering
\includegraphics[width=4.25cm,height=5cm]{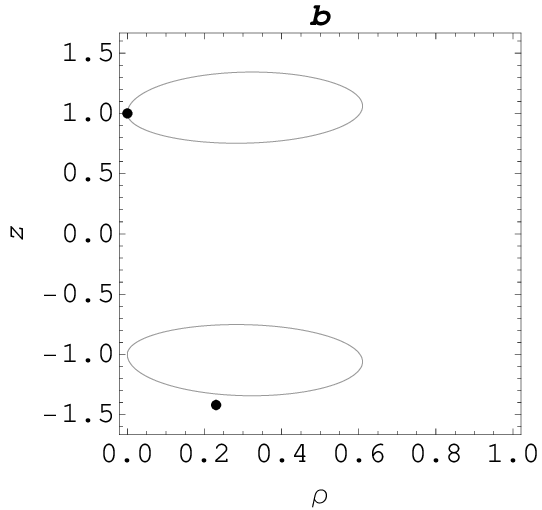}
\end{minipage}
\caption{(a) The SLS ($f=0$) in the extreme limit case and their corresponding RS for the values $M_{1}=-1$, $M_{2}=-1.1$, $Q_{1}=0.1$, $Q_{2}=-0.191$, $J_{1}=-0.148$, $J_{2}=0.179$, and $R=2.4$. The RS are located at $\rho\simeq 0.72$, $z\simeq 1.0$ and $\rho\simeq 0.74$, $z\simeq -1.07$. (b) If one of the masses is negative the system becomes corotating, for $M_{1}=1$, $M_{2}=-0.6$, $Q_{1}=0.7$, $Q_{2}=-0.1$, $J_{1}=0.306$, $J_{2}=0.855$, and $R=2$. The RS is located at $\rho\simeq 0.23$, $z\simeq -1.42$.}
\label{SLSs}\end{figure}

It has been shown for identical constituents that the presence of the electromagnetic field displaces the RS to the right side of its corresponding stationary limit surface (SLS) \cite{CLLM,ILLM}. Nevertheless, in the unequal case the RS not only is displaced to the right side, it also can move downward as shown in Fig.\ \ref{SLSs}; apparently this is due to the difference between the values of both electric charges. We mention also that if one of the masses is negative, the system becomes corotating (see Table \ref{table1}).

\vspace{-0.4cm}
\subsection{Singular surfaces}
\vspace{-0.4cm}
On the other hand, the easiest analytical proof on the regularity of the solution can be performed in the DRN sector, since the curves defined by Eq.\ (\ref{polynomials}) are now reduced in two subfamilies \cite{IMR}. The first one of them is the Majumdar-Papapetrou (MP) solution \cite{Majumdar,Papapetrou}, whose masses and electric charges are related by $Q_{k}=\pm M_{k}$ and both charges have the same sign. The curves given by Eq.\ (\ref{polynomials}) are reduced to the equation of the hyperbola,
\bea \begin{split} F_{R}&\equiv F_{MP}= \left(x+\frac{M_{1}+M_{2}}{2\alpha}\right)^{2}-\left(y-\frac{M_{1}-M_{2}}{2\alpha}\right)^{2}\\
&-\frac{M_{1}M_{2}}{\alpha^{2}}=0, \label{hyperbola} \end{split}\eea

\noi which contains two asymptotes described by
\be y= \pm \left(x + \frac{M_{1}+M_{2}}{2\alpha}\right)+\frac{M_{1}-M_{2}}{2\alpha}. \ee

\begin{figure}[ht]
\begin{minipage}{0.49\linewidth}
\centering
\includegraphics[width=4.25cm,height=5.0cm]{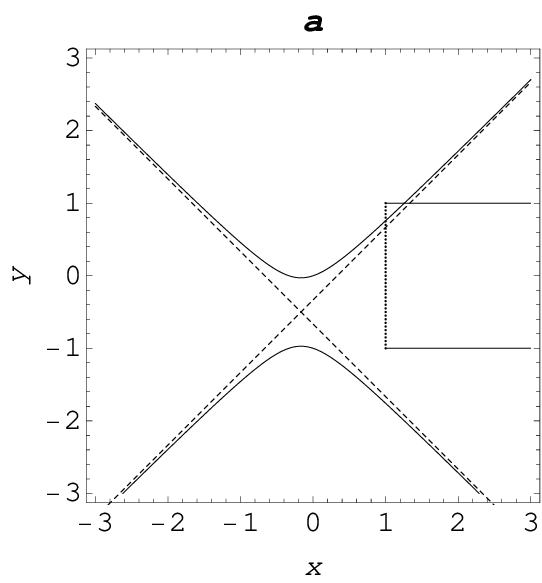}
\end{minipage}
\begin{minipage}{0.49\linewidth}
\centering
\includegraphics[width=4.25cm,height=5.0cm]{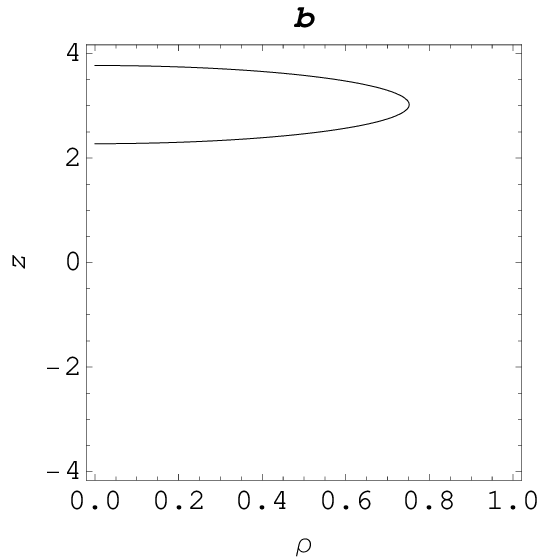}
\end{minipage}
\caption{(a) Crossing inside the region $x>1$, $|y|<1$, if one of the masses is negative (in this case $M_{1}<0$), for the values $\alpha=3$, $M_{1}=-1$, and $M_{2}=2$. (b) Appearance of the corresponding SS if $M_{1}<0$ in the MP sector.}
\label{SSMP}\end{figure}

In this respect, the conditions $x=1$ and $|y|<1$ are enough to prove that at least one asymptote is crossing inside the region $x>1$, $|y|<1$; therefore, naked sin\-gu\-la\-ri\-ties arise as SS. Let us suppose that the straight line with positive slope crosses this region, but the other one with negative slope does not (see Fig.\ \ref{SSMP}), and then we get
\bea \begin{split}  &+ \left( 1 + \frac{M_{1}+M_{2}}{2\alpha}\right)+
\frac{M_{1}-M_{2}}{2\alpha}<1,\quad \Rightarrow \quad M_{1}<0, \\
& -\left(1 + \frac{M_{1}+M_{2}}{2\alpha}\right)+\frac{M_{1}-M_{2}}{2\alpha}<-1,\quad \Rightarrow \quad M_{2}>0. \label{proofMP} \end{split}\eea

Besides, if the electric charges have opposite sign, a second electrostatic subfamily can be also obtained from Eq.\ (\ref{extreme}) where the charges and masses are related by means of
\bea \begin{split} Q_{1}&=\epsilon M_{1} \sqrt{\frac{(2\alpha
+M_{2})^{2}-M_{1}^{2}}{(2\alpha-M_{2})^{2}-M_{1}^{2}}}, \\
Q_{2}&= -\epsilon M_{2} \sqrt{\frac{(2\alpha+M_{1})^{2}-M_{2}^{2}}{(2\alpha-M_{1})^{2}-M_{2}^{2}}}, \qquad \epsilon=\pm 1, \label{chargesB} \end{split}\eea

\noi and it follows that electric charges are greater than their corresponding positive masses, according to $|Q_{k}|>M_{k}$ \cite{IMR}. As was already proved in Ref.\ \cite{IMR}, it represents the electrostatic analogue of the well-known Kerr-NUT spacetime \cite{Demianski} obtainable via Bonnor's procedure of a complex continuation of the parameters \cite{Bonnor}, whereby the sub\-fa\-mi\-ly is called Bonnor's solution (BS) \cite{Bonnor1}. In BS,
Eq.\ (\ref{polynomials}) reduces to
\bea \begin{split} F_{R}&\equiv F_{B}\\
&=D^{2}\left(x+\frac{M_{1}+M_{2}}{2\alpha}\right)^{2}-\left(y-\frac{M_{1}-M_{2}}{2\alpha}\right)^{2}=0,& \\ D&:=\sqrt{\frac{4\alpha^{2}-(M_{1}-M_{2})^{2}}{4\alpha^{2}-(M_{1}+M_{2})^{2}}}, \label{lines} \end{split}\eea

\noi which represents the geometric locus of two straight lines intersecting at an angle of $\theta=2 \arctan D$, where the straight lines are given by
\be y= \pm D \left( x + \frac{M_{1}+M_{2}}{2\alpha}\right)+ \frac{M_{1}-M_{2}}{2\alpha}. \ee

\begin{figure}[ht]
\begin{minipage}{0.49\linewidth}
\centering
\includegraphics[width=4.25cm,height=5cm]{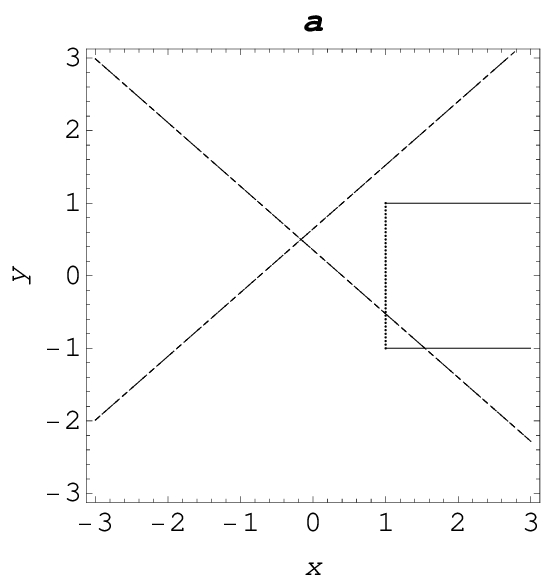}
\end{minipage}
\begin{minipage}{0.49\linewidth}
\centering
\includegraphics[width=4.25cm,height=5cm]{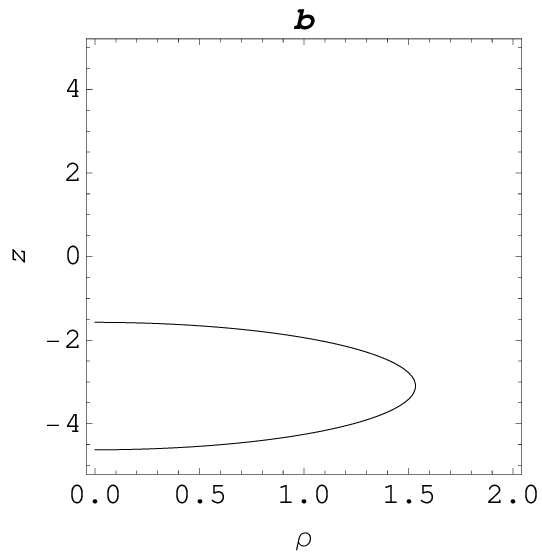}
\end{minipage}
\caption{(a) Crossing inside the region $x>1$, $|y|<1$, due to $M_{2}<0$, for the values $\alpha=3$, $M_{1}=2$, and $M_{2}=-1$. (b) Emergence of the corresponding SS if $M_{2}<0$ in Bonnor's sector.}
\label{SSB}\end{figure}

Let us suppose now that the straight line with positive slope is not crossing inside the region (the one a\-sso\-cia\-ted with $M_{1}$) while the other one does (see Fig.\ \ref{SSB}). Similarly to the MP case, we have
\begin{widetext}
\bea \begin{split}  &+ D\left( 1 + \frac{M_{1}+M_{2}}{2\alpha}\right)+\frac{M_{1}-M_{2}}{2\alpha}>1,
\quad \Rightarrow \quad \sqrt{\frac{(2\alpha+M_{1})^{2}-M_{2}^{2}}{(2\alpha-M_{1})^{2}-M_{2}^{2}}}>1, \quad \Rightarrow \quad M_{1}>0, \\
& -D\left( 1 + \frac{M_{1}+M_{2}}{2\alpha}\right)+\frac{M_{1}-M_{2}}{2\alpha}>-1,\quad \Rightarrow \quad \sqrt{\frac{(2\alpha+M_{2})^{2}-M_{1}^{2}}{(2\alpha-M_{2})^{2}-M_{1}^{2}}}<1, \quad \Rightarrow \quad M_{2}<0. \label{proofB} \end{split}\eea
\end{widetext}

\vspace{-0.2cm}

Therefore, one concludes that SS appear as a consequence of the presence of negative masses in the solution even if the total ADM mass is positive. Finally, it should be pointed out that the values for the individual masses should fulfill the Smarr mass formula $M_{k}\geq 2\Omega_{k} J_{k}+\Phi_{k}^{H}Q_{k}>0$, in order to avoid the a\-ppea\-ran\-ce of naked singularities off the axis in the solutions given by Eqs.\ (\ref{ERNST0}) and (\ref{extreme}), in agreement with the po\-si\-ti\-ve mass theorem \cite{SchoenYau1,SchoenYau2}.

\vspace{-0.2cm}
\section{Concluding remarks}
\vspace{-0.4cm}
In this paper, we developed a 6-parametric exact solution which is used as a basis to describe a binary system
of KN black hole sources separated by a massless strut, in which the seven physical Komar parameters are related by means of a simple algebraic equation. This model generalizes the known DRN problem \cite{Varzugin1,Alekseev,MankoDRN} and the double-Kerr problem \cite{ICM}. It is worthwhile to stress the fact that the paper uses the idea provided in Ref.\ \cite{MRS} to remove first the individual magnetic charges with the purpose of solving analytically the axis conditions given in Refs.\ \cite{CLLM,ILLM}. Since our model contains unequal cons\-ti\-tuents, we provide also a physical parametrization for the two half-length parameters $\sigma_{k}$ associated with the black hole horizons which characterize the solution in a more transparent way. Using the advantage that provides the physical parametrization, we show that our model saturates the Gabach Clement inequality for interacting black holes with struts \cite{Maria}. The extreme limit case is also derived by using Perj\'es' representation, and, later on, we give a numerical analysis on the presence of ring singularities off the axis and their location depending on the values of the electric charges. The numerical analysis reveals that the ring singularity located outside the ergosurface moves downward depending on the difference between values of the electric charges. Furthermore, we give a simple a\-na\-ly\-ti\-cal proof in the DRN problem on the conditions to avoid the appearance of singular surfaces, which in our opinion can be used as a complement of the well-known positive mass theorem \cite{SchoenYau1,SchoenYau2}.

It should be pointed out that the binary model could include the magnetic charges (after their elimination), but it is not trivial to derive in a simple manner the procedure (duality rotation) which allows us to include individual magnetic charges in the unequal charged case. For instance, as a particular case, if the electric charges are equal in magnitude but opposite in sign according to $Q_{1}=-Q_{2}=Q_{E}$, the electric and magnetic dipole moment can be combined in such way that the electric charge $Q_{E}$ enters in the solution by means of the transformations  $Q_{E}\rightarrow Q_{E}+ i Q_{B}$ and $Q_{E}^{2}\rightarrow |Q_{E}^{2} + Q_{B}^{2}|$. Then, the corresponding formulas for $\sigma_{k}$ are
\begin{widetext}
\bea \begin{split}
\sigma_{1}&= \sqrt{M_{1}^{2}-\left[ |Q_{E}^{2}+Q_{B}^{2}|+
\frac{J_{1}^{2}\left[\left((R+M_{1})^{2}-M_{2}^{2}\right)\left((R+M_{2})^{2}-M_{1}^{2}\right)+4
|Q_{E}^{2}+Q_{B}^{2}|R^{2}\right]}{\left[M_{1}\left((R+M_{2})^{2}-M_{1}^{2}\right)+|Q_{E}^{2}+Q_{B}^{2}|R\right]^{2}}
\right]\frac{R-M_{1}-M_{2}}{R+M_{1}+M_{2}}},\\
\sigma_{2}&= \sqrt{M_{2}^{2}-\left[ |Q_{E}^{2}+Q_{B}^{2}|+
\frac{J_{2}^{2}\left[\left((R+M_{1})^{2}-M_{2}^{2}\right)\left((R+M_{2})^{2}-M_{1}^{2}\right)+4
|Q_{E}^{2}+Q_{B}^{2}|R^{2}\right]}{\left[M_{2}\left((R+M_{1})^{2}-M_{2}^{2}\right)+|Q_{E}^{2}+Q_{B}^{2}|R\right]^{2}}
\right]\frac{R-M_{1}-M_{2}}{R+M_{1}+M_{2}}}.\label{sigmas2} \end{split}\eea

\noi where now Eq.\ (\ref{relationmomentum}) leads us to the following relation between the seven physical parameters:
\bea \begin{split}
M_{1}M_{2}(R+M_{1}+M_{2})\left[J_{1} + J_{2}+ R\left(\frac{J_{1}}{M_{1}}+ \frac{J_{2}}{M_{2}} \right)- M_{1}M_{2}
\left(\frac{J_{1}}{M_{1}^{2}}+\frac{J_{2}}{M_{2}^{2}}\right)\right]+
|Q_{E}^{2}+Q_{B}^{2}|(J_{1}+J_{2})R=0.
\label{relationmomentum2} \end{split}\eea

Finally, we must emphasize that the magnetic charges $Q_{B}$ introduced in this particular model contain opposite signs and are equal in magnitude, where now the \emph{new} electric and magnetic dipole moments are given by
\bea \begin{split}
\mathfrak{q}_{o}&=\left[Q_{E}-Q_{B}\left( \frac{J_{1}(R+M_{1}-M_{2})}{M_{1}[(R+M_{2})^{2}-M_{1}^{2}]+|Q_{E}^{2}+Q_{B}^{2}|R}-
\frac{J_{2}(R-M_{1}+M_{2})}{M_{2}[(R+M_{1})^{2}-M_{2}^{2}]+|Q_{E}^{2}+Q_{B}^{2}|R}\right)\right](R-M_{1}-M_{2}), &\\
\mathfrak{b}_{o}&=\left[Q_{B}+Q_{E}\left( \frac{J_{1}(R+M_{1}-M_{2})}{M_{1}[(R+M_{2})^{2}-M_{1}^{2}]+|Q_{E}^{2}+Q_{B}^{2}|R}-
\frac{J_{2}(R-M_{1}+M_{2})}{M_{2}[(R+M_{1})^{2}-M_{2}^{2}]+|Q_{E}^{2}+Q_{B}^{2}|R}\right)\right](R-M_{1}-M_{2}),& \\
\label{newdipoles} \end{split}\eea
\end{widetext}

\noi where $Q_{B}=0$ recovers the original electric and magnetic dipole moments obtainable from Eq.\ (\ref{dipoles}) after setting $Q_{1}=-Q_{2}=Q_{E}$. We hope to accomplish some extensions that include magnetic charges in the model under consideration.

\vspace{-0.4cm}
\section*{ACKNOWLEDGMENTS}
\vspace{-0.4cm}
The author thanks the referees for their valuable remarks and suggestions. This work was supported by Consejo Nacional de Ciencia y Tecnolog\'ia (CONACyT) M\'exico under the Sistema Nacional de Investigadores (SNI) program.


\begin{thebibliography}{99}
\bibitem{Reissner}{H. Reissner, Ann. Physik (Berlin) \textbf{355}, 106 (1916).}

\bibitem{Nordstrom}{G. Nordstr\"{o}m, Proc. K. Ned. Akad. Wet. \textbf{20}, 1238 (1918).}

\bibitem{Weyl} {H. Weyl, Ann. Phys. (Berlin) \textbf{359}, 117 (1917).}

\bibitem{Majumdar}{S. D. Majumdar, Phys. Rev. \textbf{72}, 390 (1947). }

\bibitem{Papapetrou}{A. Papapetrou, Proc. R. Irish Acad., Sect. A  \textbf{51}, 191 (1947).}

\bibitem{Varzugin1}{G. G. Varzugin and A. S. Chystiakov, Classical Quantum Gravity \textbf{19}, 4553 (2002).}

\bibitem{Alekseev}{G. A. Alekseev and V. A. Belinski, Phys. Rev. D \textbf{76}, 021501(R) (2007).}

\bibitem{MankoDRN}{V. S. Manko, Phys. Rev. D \textbf{76} 124032 (2007).}

\bibitem{BachWeyl}{R. Bach and H. Weyl, Math. Z. \textbf{13}, 134 (1922).}

\bibitem{Israel}{W. Israel, Phys. Rev. D \textbf{15}, 935 (1977).}

\bibitem{KramerNeugebauer}{D. Kramer and G. Neugebauer, Phys. Lett. A \textbf{75}, 259 (1980).}

\bibitem{IMR}{I. Cabrera-Munguia, V. S. Manko, and E. Ruiz, Gen. Relativ. Gravit. \textbf{43}, 1593 (2011).}

\bibitem{RICM}{I. Cabrera-Munguia and A. Mac\'ias, AIP Conf. Proc. \textbf{1577}, 213 (2014).}

\bibitem{DietzHoenselaers}{W. Dietz and C. Hoenselaers, Ann. Phys. (N.Y.) \textbf{165}, 319 (1985).}

\bibitem{MRS1}{V. S. Manko, E. Ruiz, and J. D. Sanabria-G\'omez, Classical Quantum Gravity \textbf{17},
3881 (2000).}

\bibitem{Hennig}{G. Neugebauer and J. Hennig, Gen. Relativ. Gravit. \textbf{41}, 2113 (2009).}

\bibitem{Newman}{E. Newman, E. Couch, K Chinnapared, A. Exton, A. Prakash, and R. Torrence, J. Math. Phys. \textbf{6}, 918 (1965).}

\bibitem{Parker}{L. Parker, R. Rufinni, and D. Wilkins, Phys. Rev. D \textbf{7}, 2874 (1973).}

\bibitem{PIW1}{Z. Perj\'es, Phys. Rev. Lett. \textbf{27}, 1668 (1971).}

\bibitem{PIW2}{W. Israel and G. A. Wilson, J. Math. Phys. (N.Y.) \textbf{13}, 865 (1972).}

\bibitem{MRR}{V. S. Manko, R. I. Rabad\'an, and E. Ruiz, Classical Quantum Gravity  \textbf{30}, 145005 (2013).}

\bibitem{Komar}{A. Komar, Phys. Rev. \textbf{113}, 934 (1959).}

\bibitem{Emparan}{R. Emparan, Phys. Rev. D \textbf{61}, 104009 (2000).}

\bibitem{EmparanTeo}{R. Emparan and E. Teo, Nucl. Phys. \textbf{B610}, 190 (2001).}

\bibitem{MRS}{V. S. Manko, R. I. Rabad\'an, and J. D. Sanabria-Gomez,  Phys. Rev. D \textbf{89}, 064049 (2014).}

\bibitem{CLLM}{I. Cabrera-Munguia, C. L\"{a}mmerzahl, L. A. L\'opez, and A. Mac\'ias, Phys. Rev. D
\textbf{90}, 024013 (2014).}

\bibitem{ILLM}{I. Cabrera-Munguia, C. L\"{a}mmerzahl, L. A. L\'opez, and A. Mac\'ias, Phys. Rev. D
\textbf{88}, 084062 (2013). }

\bibitem{Schwinger}{J. S. Schwinger, Science \textbf{165}, 757 (1969).}

\bibitem{Smarr}{L. Smarr, Phys. Rev. Lett. \textbf{30}, 71 (1973).}

\bibitem{Tomimatsu}{A. Tomimatsu, Prog. Theor. Phys. \textbf{72}, 73 (1984).}

\bibitem{Dain}{S. Dain and O. E. Ortiz,  Phys. Rev. D \textbf{80}, 024045 (2009).}

\bibitem{Perjes}{Z. Perj\'es, J. Math. Phys. (N.Y.) \textbf{30}, 2197 (1989).}

\bibitem{Maria}{M. E. Gabach Clement, Classical Quantum Gravity \textbf{29}, 165008 (2012)}

\bibitem{SchoenYau1}{R. Schoen and S.-T. Yau, Commun. Math. Phys. \textbf{65}, 45 (1979).}

\bibitem{SchoenYau2}{R. Schoen and S.-T. Yau, Commun. Math. Phys. \textbf{79}, 231 (1981).}

\bibitem{Ernst}{F. J. Ernst, Phys. Rev. \textbf{168}, 1415 (1968).}

\bibitem{Sibgatullin}{N. R. Sibgatullin, \emph{Oscillations and Waves in Strong Gravitational and
Electromagnetic Fields} (Springer-Verlag, Berlin, 1991). }

\bibitem{RMJ}{E. Ruiz, V. S. Manko, and J. Mart\'in, Phys. Rev. D \textbf{51}, 4192 (1995).}

\bibitem{Papapetrou0}{A. Papapetrou, Ann. Phys. (Berlin) \textbf{447}, 309 (1953).}

\bibitem{Simon}{W. Simon, J. Math. Phys. \textbf{25}, 1035 (1984).}

\bibitem{NUT}{E. Newman, L. Tamburino, and T. Unti,  J. Math. Phys. (N.Y.) \textbf{4}, 915 (1963).}

\bibitem{FN}{There is a misprint in formula (7) of Ref.\ \cite{ILLM}; a bar over the term $\mathfrak{a}$ is missed.}

\bibitem{ICM}{I. Cabrera-Munguia, C. L\"{a}mmerzahl, and A. Mac\'ias, Classical Quantum Gravity \textbf{30},
175020 (2013).}

\bibitem{Kinnersley}{W. Kinnersley, J. Math. Phys. (N.Y.) \textbf{18}, 1529 (1977).}

\bibitem{Weinstein}{G. Weinstein, Commun. Pure Appl. Math. \textbf{43}, 903 (1990).}

\bibitem{Carter}{B. Carter, in \emph{Black Holes}, edited by C. DeWitt and B.S. DeWitt (Gordon and
Breach, New York, 1973), p. 57.}

\bibitem{TS}{A. Tomimatsu and H. Sato, Phys. Rev. Lett. \textbf{29}, 1344 (1972); Prog. Theor. Phys. \textbf{50}, 95 (1973).}

\bibitem{ADM}{R. Arnowitt, S. Deser, and C. W. Misner, Phys. Rev. \textbf{122}, 997 (1961).}

\bibitem{Demianski}{M. Demia\'nski and E. T. Newman, Bull. Acad. Polon. Sci. Ser. Math. Astron. Phys.
\textbf{14}, 653 (1966). }

\bibitem{Bonnor}{W. B. Bonnor, Z. Phys. \textbf{161}, 439 (1961).}

\bibitem{Bonnor1}{W. B. Bonnor, J. Phys. A \textbf{12}, 853 (1979).}

\end{thebibliography}
\end{document}